# Calibración radiométrica *in-situ* de sensores satelitales de observación de la Tierra utilizando un espectroradiómetro

# Radiometric *in-situ* calibration of satelital sensors of Earth observation using a spectroradiometer

Delgado-Correal C.[1,2], García J. E.[2]

[1] Observatorio Astronómico Nacional, Universidad Nacional de Colombia, Bogotá, Colombia.
[2] Grupo de Física Aplicada y Desarrollo Tecnologico, Centro Internacional de Física-CIF, Bogotá, Colombia.

**Resumen**

Se determinó para una región del territorio colombiano, la atenuación radiativa total de la atmósfera utilizando datos satelitales de observación de la Tierra descargados por una estación terrena construida en el país, y mediciones de la reflectancía del suelo. Para ello fue necesario en primera instancia utilizar la teoría de Fourier que describe los filtros ideales de señales para hallar las funciones de transferencia entre las respuestas espectrales instrumentales de un espectroradiómetro y del sensor satelital, al cual vamos a calibrar su señal radiativa. Después teniendo en cuenta la curva de reflectancía tomada con el espectroradiómetro del suelo de una región determinada, y la información en números digitales (DN) de un píxel de la imagen satelital tomada de la misma región en el mismo lapso de tiempo, y haciendo uso de nuevo de la teoría de filtros ideales hallamos la función de transferencia entre la respuesta de ambos sensores a la radiancía del suelo. La relación entre ambas señales nos proporciona la atenuación radiativa en intensidad total de la atmósfera por píxel, la cual es fundamental para calibrar radiométricamente toda la imagen. Se obtuvo un factor de atenuación causa de la presencia de la atmósfera de la intensidad total de la radiación ($\lambda$ entre los 430nm hasta los 830nm) proveniente del suelo (de la superficie que cubre el píxel escogido) de: $H_{Total}(\lambda)_{Atmósfera} = 1{,}435 x 10^{-3}$.

**Palabras claves:** Espectroradiómetro, Satélite NOAA 18, sensor AHVRR, Función de transferencia.

**Abstract**

By using the satelital information of Earth observation unloaded by a station terrena constructed in the country and reflectances measurements of the soil, it found the total radiation attenuation of the atmosphere for a small region of the Colombian territory. It was necessary to use the Fourier's theory that describes the ideal filters of signs to find the transfer functions between the spectral response of an spectroradiometer and the satelital sensor, to which, we are going to calibrate it radiative sign. After that, we used of the reflectance spectrum of the soil taked with our spectroradiometer, the information in digital numbers (DN) of a pixel of the satelital image of the same region at the same time, and using again the theory of ideal filters we found the transfer function between the response of both sensors to the radiance of the soil. The relation between both signs provides us the total intensity of the radiation attenuation of the atmosphere for pixel, which is fundamental to do a radiometric calibration of the whole image. We found a factor of attenuation atmosphere of the radiation ($\lambda$ between 430nm to 830nm) come from the soil (of the pixel surface) of $H_{Total}(\lambda)_{Atmospehere} = 1{,}435 x 10^{-3}$.

**Introducción**

Desde el momento que una imagen satelital es descargada utilizando una estación terrena es necesario contar con instrumentación adecuada para calibrar estas imágenes radiométricamente, es decir que los valores de radiación por píxel de la imagen en cada longitud de onda que registran los sensores pasivos de percepción remota del satélite correspondan a los valores reales de radiación emitida por la superficie. Por esta razón es necesario generar rutinas de calibración radiométrica de los sensores satelitales, puesto que sí no se hace la adecuada calibración y se elimina los efectos de atenuación radiativa en la atmósfera podemos perder en algunos casos información valiosa de nuestras imágenes.

**Medida in-situ de la reflectancía del suelo utilizando un espectroradiómetro**

El dispositivo que registra la reflectancía del suelo llamado espectroradiómetro permite colectar la radiación proveniente de la tierra en las bandas del visible e infrarrojo cercano, presentando un registro continuo de la radiación emitida por la superficie en un ancho de banda espectral promedio entre los 300nm hasta los 1000nm [1].

El ancho de escena por así decirlo de un espectroradiómetro es del orden de centímetros y para obtener la radiancía un situ de una región de 1 metro (resolución típica de las bandas espectrales de los sensores del satélite IKONOS [2]) es necesario recorrer esta región y tomar las mediciones de la radiación (por longitud de onda) del suelo punto a punto a lo largo y ancho de nuestra región de interés y al final construir una grilla espacial con los valores obtenidos en cada medición. Esta medición debe ser realizada en el menor tiempo posible para considerar que en todas las mediciones tenemos las mismas condiciones atmosféricas.

**Relación entre la señal recibida por un espectroradiómetro y un sensor pasivo satelital de observación de la Tierra.**

Lastimosamente la información que registran los sensores de percepción remota no corresponde exactamente a la radiación emitida por el suelo, puesto que los gases constituyentes de la atmósfera dependiendo de la banda, absorben parte de su radiación. Por esta razón utilizar el espectroradiómetro para realizar calibraciones radiométricas *in-situ* de las imágenes satelitales se convierte en una gran herramienta para tal fin, puesto que este dispositivo registra la radiación que proviene directamente de la superficie terrestre.

A continuación se va a describir con detalle la rutina realizada para hacer calibración radiométrica in-situ de imágenes satelitales utilizando un espectroradiómetro, el cual debe encontrarse calibrado en longitud de onda y en intensidad al momento de empezar a tomar los espectros de reflectancía del suelo.

Para diseñar esta rutina de calibración se utilizó una imagen registrada del terrirorio colombiano por el sensor AHVRR del satélite NOAA-18 en la banda del visible descargada en el mes de marzo de 2009 con la estación terrena construida por el grupo de física aplicada y D.T. del CIF y un espectro de reflectancía del pasto contiguo a las instalaciones del CIF tomado en la misma fecha utilizando un espectroradiómetro.

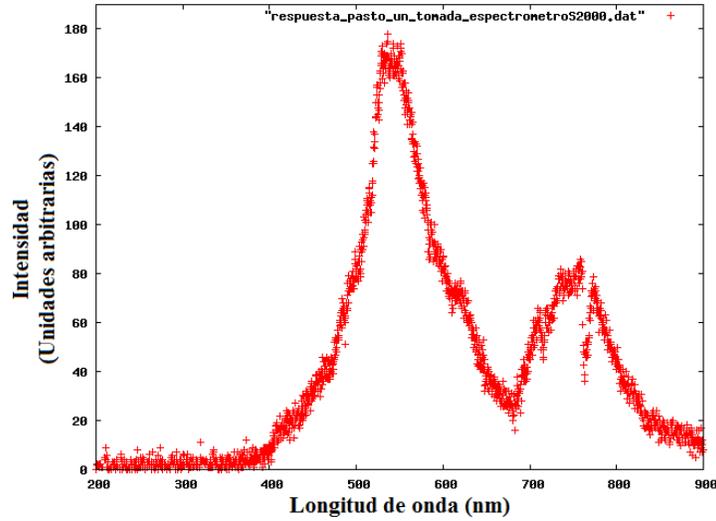

**Fig. 1.** Espectro de la radiación emitida por el pasto contiguo a la Universidad Nacional-Bogotá y registrada por un espectroradiómetro Ref.S2000.

Según la teoría de Fourier de filtros ideales tenemos que la transformada de Fourier $F(\omega)_{Entrada}$ de una señal $f(t)_{Entrada}$ que entra por un sistema (que se puede modelar como un filtro ideal con transformada de Fourier $H(\omega)_{Filtro\ Ideal}$) está relacionado con la transformada de Fourier $F(\omega)_{Salida}$ de la señal de salida del sistema de la siguiente manera [3],

$$F(\omega)_{Salida} = F(\omega)_{Entrada}\ H(\omega)_{Sistema} \quad (1)$$

En el caso de que la señal de entrada al sistema de atenuación lineal sea un haz de radiación electromagnética de cierta longitud de onda $\lambda$, entonces teniendo en cuenta que

$c = \lambda v = \lambda \frac{\omega}{2\pi}$ podemos considerar sin pérdida de generalidad que la siguiente relación es válida,

$$F(\lambda)_{Salida} = F(\lambda)_{Entrada}\ H(\lambda)_{Sistema} \quad (2)$$

Por consiguiente se satisface: $H(\lambda)_{Sistema} = \dfrac{F(\lambda)_{Salida}}{F(\lambda)_{Entrada}} \quad (3)$

$$\rightarrow H(\lambda)_{Instrumental} = \frac{F(\lambda)_{Respuesta\ Espectral\ Instrumental\ AVHRR}}{F(\lambda)_{Respuesta\ Espectral\ Instrumental\ Espectroradiómetro}} \quad (4)$$

La primera tarea de nuestra rutina de calibración es hallar la función de transferencia entre las respuestas instrumentales de ambos sensores. Para ello utilizamos la ecuación (4) donde $H(\lambda)_{Instrumental}$ representa la función deseada. Observando de nuevo (4) nos damos cuenta que es necesario conocer las respuestas espectrales instrumentales como función de longitud de onda de ambos instrumentos.

Como primera instancia es necesario obtener la información de la respuesta instrumental del sensor satelital frente a la intensidad de la radiación emitida por sus lámparas de calibración, es decir la respuesta espectral instrumental del sensor satelital que en este caso específico es denotada como $F(\lambda)_{Respuesta\ Espectral\ Instrumental\ AVHRR}$ [4][5]. Esta notación es debida a que el sensor satelital que se caracterizo fue el sensor AVHRR del satélite NOAA18 en la banda del visible.

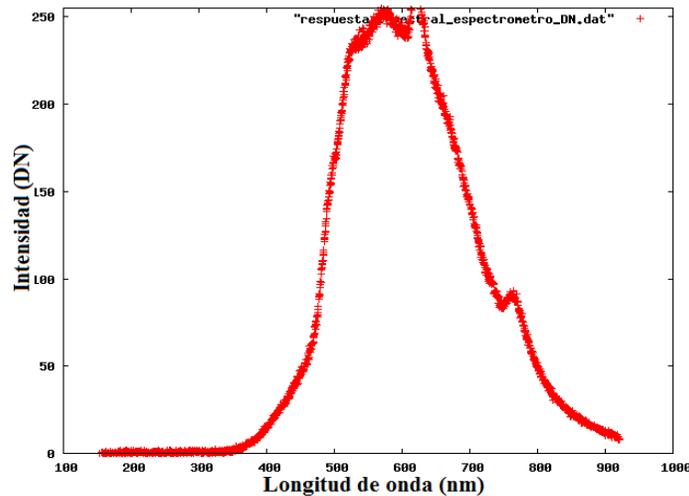

**Fig.2.** Respuesta espectral instrumental del espectroradiómetro utilizado en esta investigación (Ref. S2000).

Ahora el segundo paso para hallar $H(\lambda)_{Instrumental}$ es conocer el valor de la función espectral instrumental del espectroradiómetro utilizado en esta investigación, es decir el valor de $F(\lambda)_{Respuesta\ Espectral\ Instrumental\ Espectroradiómetro}$ por longitud de onda. Para utilizamos una lámpara de intensidad conocida para aplicar la teoría de filtros ideales y así obtener su respuesta espectral instrumental que en el caso de nuestro espectroradiómetro se logró generar el resultado que se muestra en la figura 2, en la cual podemos observar que se presentó la respuesta instrumental en números digitales (DN) (a 8bits, es decir 0-255) puesto que al aplicar la teoría de filtros ideales es más conveniente utilizar esta unidad debido a que el valor de radiancía por píxel de las imágenes satelitales generalmente se expresa en esta unidad.

Finalmente remplazando cada valor de intensidad por longitud de onda de $F(\lambda)_{Respuesta\ Espectral\ Instrumental\ Espectroradiómetro}$ y $F(\lambda)_{Respuesta\ Espectral\ Instrumental\ AVHRR}$ en la ecuación (3) y, después graficando cada valor de $H(\lambda)_{Instrumental}$ por longitud de onda, obtenemos la función de transferencia instrumental entre ambos sensores, la cual se muestra en la figura 3.

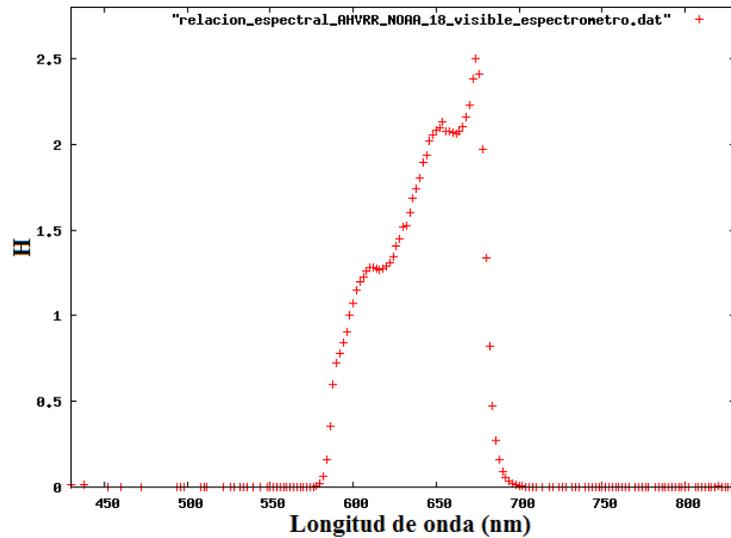

**Fig .3.** Función de transferencia instrumental entre el sensor AVHRR del satélite NOAA18 y el espectroradiómetro utilizado en esta investigación.

La segunda tarea de nuestra rutina de calibración es simular la respuesta radiativa del sensor satelital frente a la radiación emitida por el suelo como sí el sensor estuviera a unos pocos metros sobre la superficie terrestre. Para ello utilizamos la siguiente relación teniendo en cuenta el análisis hecho para obtener (2),

$$F(\lambda)_{AVHRR\ Simulada} = F(\lambda)_{Medida\ Espectroradiómetro}\ H(\lambda)_{Instrumental} \quad (5)$$

Donde $F(\lambda)_{Medida\ Espectroradiómetro}$ representa la medición por longitud usando el espectroradiómetro de la radiación liberada por el pasto contiguo a las oficinas del CIF y $H(\lambda)_{Instrumental}$ es la función de transferencia hallada anteriormente.

Después de remplazar los valores correspondientes en la ecuación (5) obtenemos la respuesta simulada en la cercanía de la superficie (1 metro) del sensor AVHRR frente a la radiación emitida por el pasto contiguo al CIF (Ver fig. 4).

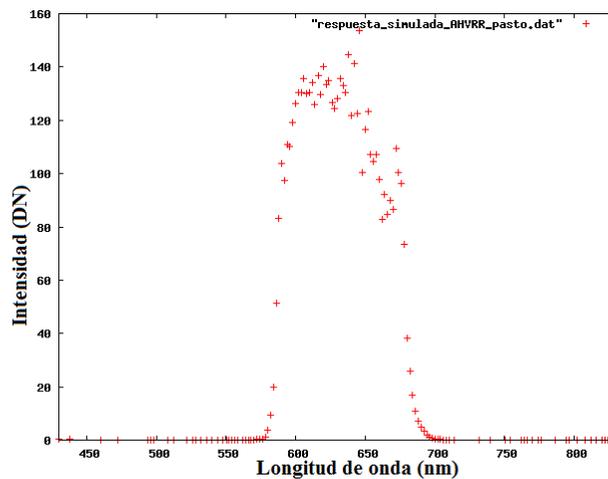

**Fig.4.** Respuesta espectral simulada a 1 metro de la superficie del sensor AVHRR del satélite NOAA18 a la radiación emitida por el pasto contiguo al CIF.

La tercera tarea de nuestra rutina de calibración es hallar la atenuación radiativa en intensidad total de la atmósfera. Entonces para comparar la radiación recibida por el sensor real a 854km [8] y la recibida por el sensor simulado (utilizando el espectroradiómetro) a 1 metro debemos asegurar tener en ambas mediciones la misma condición de iluminación, es decir que sean hechas en un lapso de tiempo similar. Además debemos tener en cuenta la posición orbital del satélite, es decir que observando el satélite en coordenadas horizontales se debe tener en cuenta el ángulo cenital $\theta$ en el momento que queramos aplicar de nuevo la teoría de filtros ideales para comparar ambas respuestas, donde se considera a la atmósfera como el sistema filtro de la señal radiativa que proviene de la superficie terrestre. En este tratamiento se considera que el sensor simulado registra la información real de la radiancía del suelo, es decir que según lo observado en (3) la información que registra este sensor entra en (3) como la $F(\lambda)_{Entrada}$ y, la radiancía que registra el sensor real es tomada en (3) como la $F(\lambda)_{salida}$.

Así la siguiente relación empieza a tener mucho sentido para nuestros propósitos iniciales,

$$H_{Total}(\lambda)_{Atmosfera} = \frac{F_{Total}(\lambda)_{Sensor\ Real}\cos(\theta)}{F_{Total}(\lambda)_{Sensor\ Simulado}} \quad (6)$$

Numéricamente para empezar a realizar este análisis se utilizó la información de la radiación registrada de un píxel de una imagen descargada que correspondiera a un sitio que tuviera un promedio de vegetación constante en el área del píxel, y así se encontró que $F_{Total}(\lambda)_{SensorReal} = 8DN$. Después integrando la figura 4, es decir en términos prácticos sumando todos sus valores de intensidad por longitud de onda se obtuvo que $F_{Total}(\lambda)_{SensorSimulado} = 5575DN$. Finalmente sabiendo que para la fecha de la toma de la imagen $\theta$ fue de 22 grados se obtuvo que $H_{Total}(\lambda)_{Atmósfera} = 1,435x10^{-3}$. Donde este valor representa el factor de atenuación de la intensidad total de la radiación (entre los 430nm hasta los 830nm) proveniente del suelo (En la superficie que cubre el píxel escogido) a causa de la presencia de la atmósfera. La última tarea del proceso de calibración es hallar la radiancía real que debe haber registrado cada píxel remplazando el valor encontrado de $H_{Total}(\lambda)_{Atmósfera}$ en la ecuación (6) teniendo en cuenta los valores de radiancía de cada píxel de la imagen[1]. Así finalmente después de hacer este proceso en todos lo píxeles de la imagen obtenemos nuestra imagen corregida radiométricamente.

**Conclusiones**

Se realizó una rutina de calibración radiométrica in situ de imágenes satelitales de resolución baja utilizando un espectroradiómetro. Este procedimiento podrá ser útil para calibrar radiométricamente las imágenes que registren los sensores de percepción remota del futuro satélite colombiano de observación de la Tierra.

**Agradecimientos**

---

1   En este proceso se asumió que la atmósfera es espacialmente homogénea.